# Rotation and dissociation dynamics of a single $O_2$ molecule on the Pt(111) surface determined from a first principles study

Qiang Fu[1,2], Jinlong Yang[1], Yi Luo[1,2*]

[1] Hefei National Laboratory for Physical Science at the Microscale,
University of Science and Technology of China, Hefei, Anhui 230026, P.R. China and
[2] Department of Theoretical Chemistry, School of Biotechnology,
Royal Institute of Technology, SE-10691 Stockholm, Sweden

(Dated: October 27, 2010)

The STM induced rotation and dissociation dynamics of a single oxygen molecule on the Pt(111) surface have been finally determined by first principles calculations together with a newly developed statistical model for inelastic electron tunneling. Several long-standing puzzles associated with these dynamic processes in this classic system have been fully resolved. It is found that the unexpected low energy barrier of the $O_2$ rotation is originated from an ingenious pathway, while the prior occupation of the metastable hcp-hollow site after the $O_2$ dissociation can be attributed to a dynamic process of surface accommodation. The experimentally observed non-integer power-law dependence of the rotation rate as a function of the current can be perfectly explained by taking into account the randomness of multi-electron inelastic tunneling processes.

PACS numbers:

Rich dynamic processes of a single oxygen molecule on the Pt(111) surface were achieved a decade ago by elegant scanning tunneling microscopy (STM) experiments [1, 2]. These experiments demonstrated for the first time that the inelastic electron tunneling could be used to manipulate and control chemical reactions of a single molecule on the metal surface [1, 2]. Such a superb ability of STM has been widely employed in recent years to study single molecular chemistry of many different systems[3–5]. However, the underlying mechanisms of many dynamic processes are still largely unexplored even for the first experiments of this kind. It was observed that under the manipulation of the STM tip, the single oxygen molecule can be dissociated into two oxygen atoms [1] or rotate among three equivalent orientations [2] on the Pt(111) surface. But, the ultralow energy barrier of the $O_2$ rotation is counter-intuitive in view of its large adsorption energy on the Pt(111) surface. It is also difficult to understand the prior occupation of the metastable hcp-hollow site after the $O_2$ dissociation since the fcc-hollow site is more stable by about 0.4eV in energy. Another outstanding problem is the non-integer power-law dependence of the rotation rate as a function of the tunneling current. It is known that the exponent of this power-law relation represents the number of inelastic electrons needed for a rotation event. Non-integer exponent means, according to the conventional theory [6], that the event of a fractional electron tunneling could take place, but it is simply impossible in reality. A good solution to all these problems could be very useful for understanding the STM induced single molecular chemistry in general since the energy barrier and the reaction rate are the most fundamental parameters for all chemical processes.

In this letter, we demonstrate that with a systematic first principles study, it is possible to explain the experimental findings through identifying the most favorable rotation and dissociation pathways for a single $O_2$ on the Pt(111) surface. We find that the single oxygen molecule can rotate on the Pt(111) surface in an ingenious way that only requires very little energy to overcome the barrier. We also discover an interesting dynamic process of the surface accommodation that rationalizes the observed prior occupation of the metastable hcp-hollow site after the $O_2$ dissociation. With the application of a recently developed statistical model, the non-integer exponent of the power-law relationship between the rotation rate and the current can be fully reproduced and explained, which comes from the statistical average of all possible n-electron events. It is noted that the statistical property of the inelastic multi-electrons tunneling process holds the key for understanding the peculiar switching behavior of a single molecular switch in a recent STM experiment [7].

Spin-polarized calculations were performed with the Vienna Ab-initio Simulation Package (VASP) [8, 9]. Projector augmented wave (PAW) potentials were employed [10] within a plane wave basis set expanded up to a cut-off energy of 400eV. Exchange correlation was described by the generalized gradient approximation (GGA-PW91) [11]. GGA-PBE [12] was also used for comparison. The Pt(111) surface was modeled by a periodic supercell of four layers separated by a 12Å thick vacuum region. A rectangular $2\sqrt{3} \times 4$ unit cell was employed at the optimized equilibrium lattice constant of 3.983Å (the experimental value is 3.92Å). The brillouin zone sampling was carried out using a 4×3×1 Monkhorst-Pack meshes of k-points. During the optimizations, the uppermost two platinum layers as well as the adsorbed molecule were allowed to relax till atom forces below 0.03eV/Å. The climbing-image nudged elastic band method (c-NEB) was used to explorer the adiabatic minimal energy pathway with the minimization of a set of four images [13]. Theo-retical STM images were simulated with Tersoff-Hamann formula [14], i.e. integrating spatially resolved density of state (DOS) in energy from a bias potential to the Fermi level.

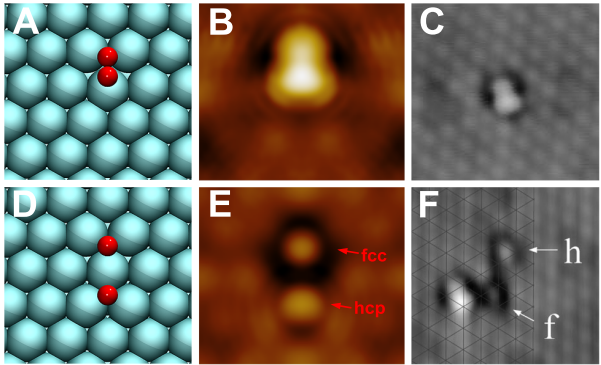

FIG. 1: Optimized structures (A and D), simulated STM images (B and E), and the STM topographic images from the experiments of Ref. [2] and [1] (C and F) for the adsorption of oxygen molecule and atoms. Copyright 1998 American Association for the Advancement of Science and 1997 American Physical Society.

We first determine the adsorption structures of an oxygen atom and a single molecule on the Pt(111) surface. This is an issue that has been extensively investigated both experimentally [1, 2, 15, 16] and theoretically [17–20]. The oxygen molecule was found in experiments to adsorb on the face centered cubic (fcc) threefold hollow sites with a lightly canted top-hollow-bridge configuration [2, 17] as shown in Fig. 1A. The adsorption energy $E_{ad}$ was estimated to be 0.79eV, which agrees well with the theoretical result of 0.68eV [17]. The corresponding STM image has also been simulated, as shown in Fig. 1B, which resembles very well with the "pear" shape observed in the experiment [2] (Fig. 1C). The asymmetric appearance of the molecule in the STM image is resulted from the tilted adsorption structure. For the adsorption of an oxygen atom, we only consider the fcc- and hcp-hollow sites, which are the ones identified in the STM experiments [1]. Our calculated adsorption energies for the fcc- and hcp- hollow sites are 4.66 and 4.24eV, respectively, which are in good agreement with the experiments [1, 16] and previous calculations [19, 20]. We also consider the co-adsorption of two oxygen atoms, *i.e.* one locates at the fcc-hollow and another on the hcp-hollow site, to mimic the final product of the dissociation (Fig. 1F). In this case, the adsorption energy is about 1.05eV more than that of the molecular adsorption (this value increases to 1.42eV when the two atoms are far away from each other). The large energy difference causes non-thermal motion of the oxygen as "hot atoms" [21], and therefore explains the fact that the two oxygen atoms can separate as far as three lattice constants after the dissociation [1]. The simulated STM image as displayed in Fig.

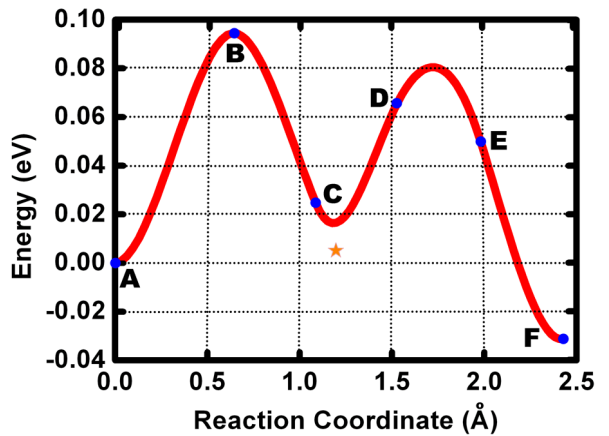

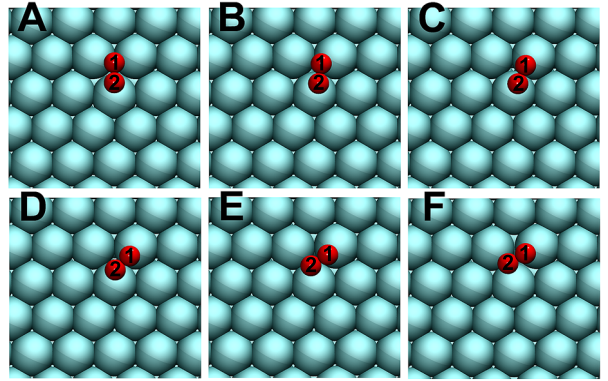

FIG. 2: Energy profile and the corresponding structures for the rotation process of a single $O_2$ molecule on the Pt(111) surface. The red line is fitted according to the energy and the force (first derivative of energy) of each images (blue dots). It should be noted that 'A' and 'F' should have the same energy, and the two barriers should have the same height because of the symmetry. The slight asymmetry of the potential energy surface is non-physically introduced by the slab model used in the simulations. The orange star denotes the exact location of the paramagnetic state.

1E fits well with the one obtained from the experiment[1] (Fig. 1F).

As we have mentioned, it was found experimentally that a single oxygen molecule can rotate reversibly among three equivalent orientations on the Pt(111) surface under the excitation of inelastic electrons [2]. In the experiment only very small energy is needed to induce the rotation [2]. It is quite surprising to see the ultra-low rotation barrier because the interaction between the oxygen molecule and the Pt(111) surface is known to be strong. Besides, the experimentally estimated value (0.15eV<$E_{rot}$<0.175eV) of the rotation barrier has not yet been verified by any theoretical calculations. In order to resolve this controversy, we have carried out systematic calculations to explore all possible pathways for the rotation process. Finally, an ingenious pathway is identified, as shown in Fig. 2. Although we consider only one rotation, the details of the reversible $O_2$ rotation among the three equivalent orientations can be revealed in view of the threefold symmetric arrangement of the Pt(111) substrate. It can be seen that during the whole process the oxygen molecule rotates 60°, and does not undergo large displacement. The small changes of bond length also indicate that the oxygen molecule holds intact during the rotation. Our calculations also confirm the ultra-low rotation barrier, and the calculated value 95meV (98meV from GGA-PBE) agrees well with the experiments [2]. As a note, it is found that the $O_2$ rotation could pass through the vicinity of a paramagnetic state with symmetric top-bridge-top configuration, that has been theoretically located before [17, 18]. The height of the two oxygen atoms is exchanged after the paramagnetic state, which seems to reminisce the H-bond donor-acceptor exchange in the water dimer diffusion process [22, 23].

The STM induced $O_2$ dissociation on the Pt(111) surface has been firmly demonstrated in experiments [1]. It was found that after the dissociation, the two oxygen atoms locate on the fcc and the hcp sites respectively, or both occupy the hcp sites. However, previous works showed that the fcc site is more stable for oxygen atom adsorption [1, 16, 19, 20] (0.44eV in our calculations). It is also worth noting that through thermal dissociation only fcc sites were occupied by the oxygen atoms [24], in sharp contrast with the phenomena here [1]. Thus, a question arises why the metastable hcp site has the priority to be occupied over the fcc one. The detailed process of $O_2$ dissociation was theoretically studied and the results are put in Fig. 3. The two oxygen atoms depart away from each other until reach an intermediate state shown in Fig. 3F. In this state one oxygen occupies a hcp-hollow site and the other locates on a top site. The top-site oxygen atom can continue to move until arrives at another hcp-hollow site through a barrierless process (Fig. 3 F-I). In addition, after the dissociation this top-site oxygen can also move further towards a fcc-hollow site through the "cannon ball" mechanism [25], because the energy gained from both STM excitation and exothermic dissociation could be distributed asymmetrically between the two atoms of $O_2$ molecule in a tilted configuration [25]. Therefore the prior occupation of the metastable hcp-hollow site can be attributed to a dynamics process of the site accommodation. An estimation of the energy barrier in the dissociation process is also obtained (0.38eV from GGA-PW91) and agrees very well with the experiments (within the range of 0.35 to 0.38eV) [1].

The $O_2$ rotation and dissociation are known to be induced by the inelastic electron excitations[1, 2, 6]. In this case, a power-law relationship between the rate of the event and the tunneling current could always be obtained. The exponent of the power-law represents the number of electrons involved in the excitations, which should in principle be an integer[1, 2, 6]. However, it was already found in the first experiment, like the

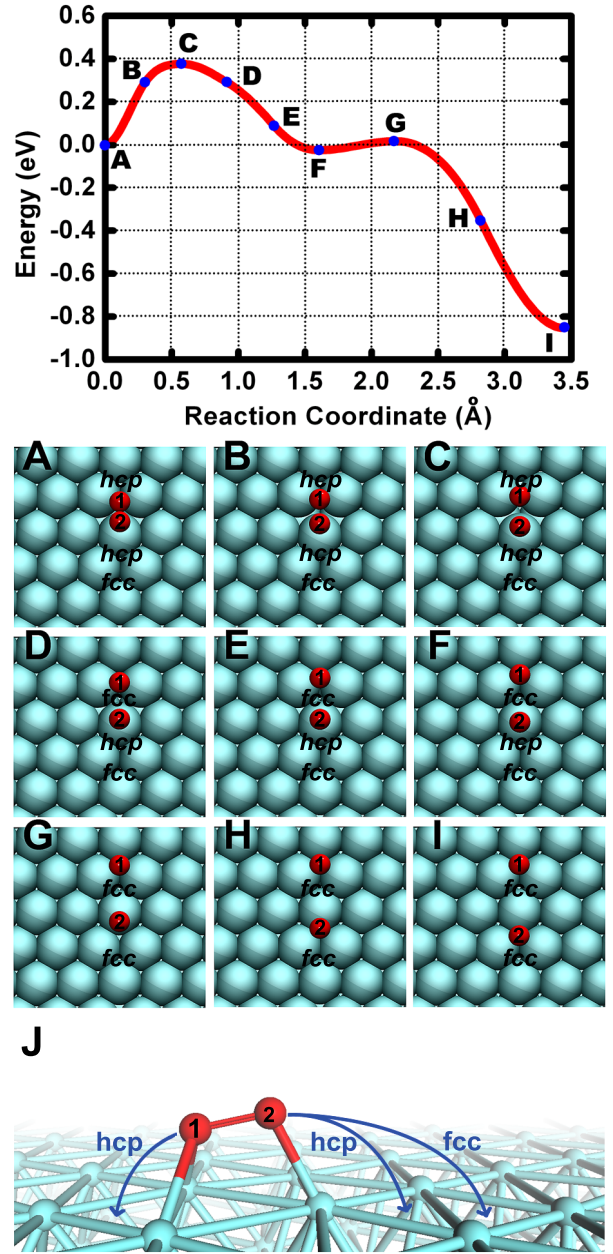

FIG. 3: Energy profile and the corresponding structures for the dissociation process of a single $O_2$ molecule on the Pt(111) surface. Schematic representation of the "cannon ball" mechanism is shown at the bottom.

$O_2$ rotation[2], that the exponent could be non-integer. Over the years there are increasing experimental evidence showing such a non-integer behavior[6, 7, 26]. Recently we put forward a statistical model that can satisfactorily explain the non-integer exponents observed for a switching process of a single molecule [7]. The basic model is schematically drawn in Fig. 4A. Under the inelastic electron excitations, the adsorbed molecule changes along the potential energy surface by accumulating the energy. It should be noted that the process involves both energy gain from the inelastic electrons and energy lose because of its interaction with the surrounding[7]. In other words, the randomness of the inelastic electron tunneling, as well as the statistical average of multi-electrons events should be taken into account. It can be seen from Fig. 4 that this statistical model can indeed reproduce the experimental results for the $O_2$ rotation process[2]. The simulated exponents ($N$) under different biases agree quite well with the experiments (Fig. 4B). Statistical distributions of various n-electron events are shown in Fig. 4. It is clear that the non-integer exponent does not correspond to one particular n-electron process, but is a statistically averaged value that comes from contributions of various n-electron events. For instance, at the bias of 0.10V, the inelastic excitation process is mainly a mixture of 47% two-electron and 38% three-electron processes. The statistical distribution of various n-electron events is originated from the competition between the energy gain and lose. Our simulations have shown that these non-integer exponents could be obtained when the the relaxation rate ($1/\tau$) is comparable to the excitation probability ($\boldsymbol{P}$).

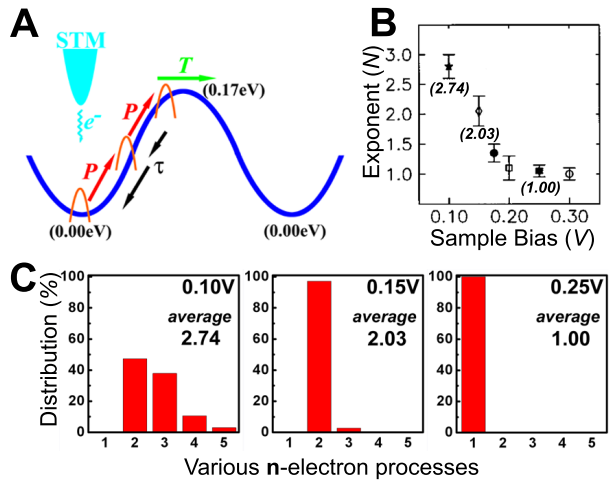

FIG. 4: (A) Schematic picture of the statistical model; (B) The observed exponent values from Ref. [2] (Copyright 1997 American Physical Society). Values in the parentheses are the simulated ones for comparison. (C) Statistical distributions of the number of inelastic electrons ($N$) under different biases. The red bars represent the percentage of the corresponding n-electron process.

It can be easily understood that once the energy gaining probability is much higher than that of energy losing, the exponent in the power-law relationship should approach an integer as observed in several experiments[27, 28], including the case of the $O_2$ dissociation [1]. By fitting the experimental results of the $O_2$ dissociation, one can see that the relaxation is too slow to influence the energy accumulation process. This observation agrees well with those of Gao et al., which found that the excitation rate is about $54$ times as fast as the rate of relaxation [1].

In summary, we have performed a systematic theoretical study on the STM induced rotation and dissociation dynamic processes of a single $O_2$ molecule on the Pt(111) surface with the-state-of-the-art computational methods. It gives satisfactory explanations for several unusual experimental phenomena that had not been fully understood for a decade. Our study shows that the current theory has reached a level of sophistication and completeness that allows for accurately describing the dynamics of a single molecule on the metal surface.

**Acknowledgement:** This work is supported by the National Key Basic Research Program of China (2010CB923300, 2006CB922004), by the Swedish Research Council (VR), and by the National Natural Science Foundation of China (Grant Nos. 20773112, 50721091, 20925311). We acknowledge the support from the Swedish National Infrastructure for Computing (SNIC), the USTC-HP HPC project, the SCCAS and the Shanghai Supercomputer Center.